\documentclass[11pt]{article}
\usepackage{spconf,amsmath,epsfig,color}
\usepackage{flushend,cite}
\usepackage[hyphens]{url}
\usepackage{upgreek}

\usepackage[utf8]{inputenc}
\usepackage[T1]{fontenc}
\usepackage{amsmath}
\usepackage{graphicx} 
\usepackage{hyperref}
\usepackage[caption=false]{subfig}

\usepackage{xcolor}
\usepackage[thickqspace,cdot,squaren]{SIunits}
\title{Landslide Geohazard Assessment with Convolutional Neural Networks using Sentinel-2 Imagery Data} % Landslide Geohazard Assessment with Deep Convolutional Neural Networks using Satellite Imagery

\name{\small{${\it S.L. Ullo}^1$, ${M. S. Langenkamp}^2$, ${T.P. Oikarinen}^2$, ${M.P. Del Rosso}^1$, ${A. Sebastianelli}^1$, ${F. Piccirillo}^1$},${S. Sica}^1$}
\address{$^{(1)}$ University of Sannio, Benevento (Italy), {\it ullo@unisannio.it, stefsica@unisannio.it,}\\
{\it mariapia.delrosso@gmail.com, alessandro.sebastianelli1995@gmail.com,}\\
{\it federica.piccirillo95@gmail.com}\\
$^{(2)}$ Massachusetts Institute of Technology (MIT), USA, {\it maxnz@mit.edu, tuomas@mit.edu }}

\begin{document}
\maketitle

\begin{abstract}
In this paper, the authors aim to combine the latest state of the art models in image recognition with the best publicly available satellite images to create a system for landslide risk mitigation. We focus first on landslide detection and further propose a similar system to be used for prediction. Such models are valuable as they could easily be scaled up to provide data for hazard evaluation, as satellite imagery becomes increasingly available.
The goal is to use satellite images and correlated data to enrich the public repository of data and guide disaster relief efforts for locating precise areas where landslides have occurred.
Different image augmentation methods are used to increase diversity in the chosen dataset and create more robust classification. The resulting outputs are then fed into variants of 3-D convolutional neural networks. A review of the current literature indicates there is no research using CNNs (Convolutional Neural Networks) and freely available satellite imagery for classifying landslide risk. The model has shown to be ultimately able to achieve a significantly better than baseline accuracy.
\end{abstract}

\vspace{-0.5cm}
\section{Introduction}
\label{sec:intro}

Landslides are an increasingly significant concern in a world of increasing climate volatility, and there have been a number of efforts to improve the predictive technology around using different methods to improve landslide monitoring techniques\cite{paper1, paper2}. 
Landslides often happen without clear warning. Consequences are catastrophic in terms of human losses. Governments are therefore interested in collaborating with researchers to detect landslides and mitigate their effects. So far, methodological investigations have been largely focused on using labor-intensive preprocessing techniques on less than half a dozen landslides \cite{CASAGLI201692}.

Authors tried to enhance the work of feature estimation    accomplished in \cite{ullo_old1} and \cite{6351884}, moving toward the involvement of  machine learning techniques, such as logistic regression and Support Vector Machines (SVM) \cite{Cortes1995},  used with good results for landslide prediction as demonstrated by a review of the relevant literature        \cite{DAS2010627,Akgun2012}, \cite{Kavzoglu2014,PENG2014287}.
However, these methods often rely on exact features of the geographic region such as elevation, gradient of the slope and soil type or extensive preprocessing of the images. A review of the current literature indicates there is no research using CNNs (Convolutional Neural Networks) and freely available satellite imagery for classifying landslide risk.
Such a model would require fewer features and less labor-intensive preprocessing of the data, yielding a system that requires lower additional effort with each set of images. 

This paper aims to create such a system for the accurate detection of landslides. The selection of image recognition model is inspired in part by Krezhevsky et al. paper \cite{NIPS2012_4824} on Convolutional Neural Networks, but also prior work using 3-dimensional convolutions, which allows the model to learn the dimension of time \cite{3D_CONV}. To the best of the authors' knowledge, combination of CNNs (Convolutional Neural Network) and public landslide satellite imagery has not been published and hence this paper serves as a novel contribution to the field of geohazard assessment.
%\vspace{-0.5cm} 

\section{use of data}

The primary focus of this paper has been on catastrophic to large landslides (as defined by the Landslide catalog) that have occurred from 2015 to 2017, due to data availability (explained below). The NASA Open Data Global Landslide Catalog \cite{catalog} was used as the starting point for landslides that fit the criteria. It contains coordinates, scale (ranging from 'catastrophic' to 'small'), date, among other information. 
Initially, both Sentinel-1 and Sentinel-2 data were analysed for usability. However, Sentinel-1 images proved to be computationally prohibitive due to the large amount of preprocessing required. Therefore, it was decided that Sentinel-2 data were better for creating a large database, necessary for training a deep learning model \cite{DL_scalability}.

Since the ESA Copernicus Sentinel-2 \cite{sentinel2_site} optical images suited our purposes most closely, and because the launch of Sentinel-2A happened in June 23, 2015, we only analyzed landslides that occurred after June, 2015. Furthermore, since the catalog ends in September, 2017, we did not look at landslides that happened after this date. The landslides were then filtered by size, focusing on the largest first, so that the model could more easily learn an internal representation of a landslide. Only landslides where the location was known with an accuracy of 1 km or less were used. Images that included noticeable cloud coverage were also discarded.
Among thousands of landslides available in the Global Landslides Catalog, twenty landslides were chosen for preliminary testing. A sample is shown in Table \ref{tab:data}.
\label{sec:case_study}

\begin{table}[!ht]
\caption{A sample of landslides of interest}
\vspace{10pt}
\label{tab:data}
\centering
\scalebox{0.45}{
    \begin{tabular}{l|l|l|l|l|l}
    \hline
    Location & Landslide date & Landslide size & Type & Latitude & Longitude\\
    \hline
    Greenland (Hill near Nuugaatsiaq)& 06/17/2017 & catastrophic & landslide & 71,53659933 & -53,20874578\\
    \hline
    California(Mud Creek) & 05/20/2017 & very large & landslide & 35,865628 & -121,43238\\
    \hline
    China (Village of Xinmo) & 06/24/2017 & catastrophic & debris flow & 32,08087401 & 103,6656168\\
    \hline
    Indonesia (Jalan Melati) & 08/19/2016 & very large & mudslide & -6,311708 & 106,801076\\
    \hline
    Colombia (Mocoa) & 03/25/2017 & very large & landslide & 1,15189804 & -76,639923\\
    \hline
    Switzerland (Piz Cengalo) & 23/08/2017 & very large & debris flow & 46,29694 & 9,595744\\
    \hline
    UK (Bridgeport and West Dorset) & 29/06/2017 & very large & landslide & 50,70838 & -2,75802\\
    \hline
    Bosnia \& Herzegovina (Kakanj) & 24/02/2017 & very large & landslide & 44,14354 & 44,14354\\
    \hline
    \end{tabular}
}
\end{table}
 \vspace{-0.5cm} 
\section{Proposed Model}
\begin{figure}
    \centering
    \includegraphics[scale=0.5]{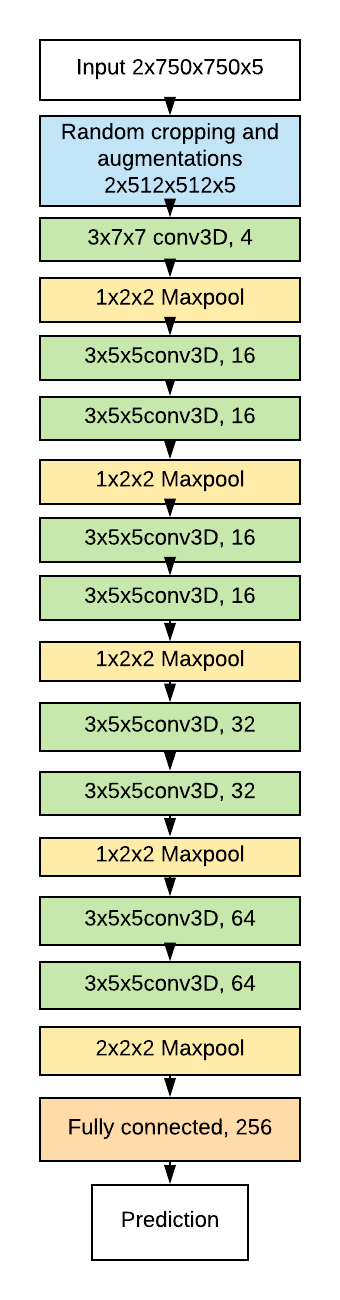}
    \caption{Diagram of model}
    \label{fig:model_diagram}
\end{figure}

The model is a convolutional neural network, which performs a set of differentiable mathematical operations on the input values, described by the model's set of weights, to produce an output. Initially the weights are randomized, and the network is then trained to minimize the loss function. The loss function in this case is the negative log-likelihood loss, described by:
\begin{equation}
    \centering
     L(X) = -\sum_{x,y\in\mathcal{X}} p(x)\cdot log(y)+(1-p(x))\cdot log(1-y)
    \label{eqn:loss_function}
\end{equation}
Where X is our training dataset of image, label pairs and p(x) is the prediction of our model for input x. 
We use the backpropagation algorithm and the Adam optimizer \cite{Adam} to repeatedly update the weights in a direction that minimizes the loss function. 
The model proposed is illustrated in Fig. \ref{fig:model_diagram}. First, it extracts remote sensing data, in the form of Sentinel-2 images and also historical weather data (rainfall, humidity, cloud cover), for preprocessing.\\ During the preprocessing stage, a single landslide image is made into multiple images, by taking a randomly sampled window of the original image that still is guaranteed to contain the landslide.

After the preprocessing stage, the images are fed into the CNN model. The CNN contains 8 learned layers. A single input example consists of two images (one image before the landslides, one image after the landslide), each of size 512 by 512 pixels, and an extra dimension of size 5 for the Sentinel-2 bands. We used Bands 2, 3, 4, 8 and 12 for the level of resolution and independent information they provide.
This forms a 2x512x512x5 input which is then processed using 3D-convolutions.
Additionally, image rotations and flipping were used to increase the number of images to train on. This contributed significantly to the model's gains in accuracy.

 \vspace{-0.5cm} 
\section{Proposed analysis}
\label{sec:results}
\begin{figure}[!ht]
  \centering
  %\begin{tabular}[b]{c}
    \includegraphics[scale=0.35]{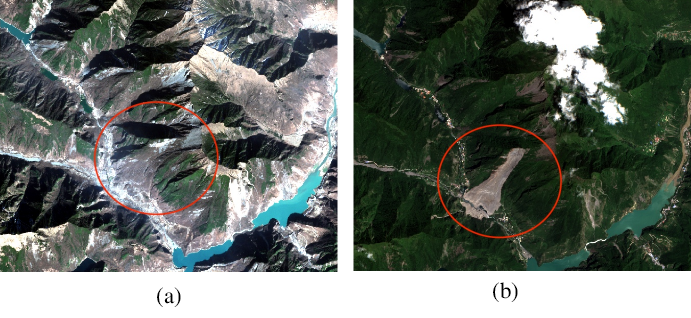}\\
    %\small (a)
  %\end{tabular} \qquad
  %\begin{tabular}[b]{c}
  %  \includegraphics[width=.85\linewidth]{images/after.jpg} \\
   % \small (b)
 % \end{tabular}
  \caption{An example of Sentinel 2 image acquired before (a) and after (b) the landslide in Xinmo, China (Table 1)}
  \label{fig:before_after}
\end{figure}

 In order to train the machine learning algorithms, we downloaded several images for each landslide (both before and after the landslide). This allows us first to create multiple pairs to train the detection model, and later to potentially use the sequence of images to train the predictive model. We also used pairings of two images before the landslide as examples where landslides did not occur. The constellation of two satellites (Sentinel-2A and Sentinel-2B) allows us to reach a 5-day geometric revisit time \cite{revisit}.
 The images were preprocessed using the Google Earth Engine developer console \cite{gee}, where the images could be filtered by cloud cover and cropped by coordinates.
An example of the completely processed images is shown in the Fig. \ref{fig:before_after}, where Sentinel-2 images acquired before and after the landslide in Xinmo (China) are shown. In the second image the landslide is clearly detectable and even if a cloud is present,   the analysis can still be done because the cloud does not cover the area of interest.
It's worth to say that the images may appear different just because taken in different light conditions. 

%\begin{figure}[!ht]
 %    \centering
 %    \subfloat[][a]{\includegraphics[scale=0.3]{images/before.jpg}\label{fig:after}}
 %    \subfloat[][b]{\includegraphics[scale=0.3]{images/after.jpg}\label{fig:before}}
 %    \caption{An example of Sentinel 2 image acquired before (a) and after (b) the landslide in Xinmo, China (see table 1)}
 %    \label{fig:after_before}
%\end{figure}

%\begin{figure}[!ht]
  %\centering
  %\begin{tabular}[b]{c}
    %\includegraphics[width=.85\linewidth]{images/before.jpg}\\
   % \small (a)
 % \end{tabular} \qquad
  %\begin{tabular}[b]{c}
  %  \includegraphics[width=.85\linewidth]{images/after.jpg} \\
  %  \small (b)
 %\end{tabular}
  %\caption{An example of Sentinel 2 image acquired before (a) and after (b) the landslide in Xinmo, China (Table 1)}
  %\label{fig:before_after}
%\end{figure}

The model was trained on 20 different landslides, using a 5-fold cross validation method \cite{kohavi1995study} to corroborate  on 4 landslides at a time. For evaluating the accuracy we decided to use balanced accuracy, which is the mean of the accuracies on examples from each class (in our case, yes landslide and no landslide). For example a network that predicts every example to be a landslide would achieve a 100\% accuracy on landslides and 0\% accuracy on images without a landslide, resulting in a balanced accuracy of 50\% regardless of the relative frequencies of the classes. We trained our network for 120 epochs with each training set and at the end an average balanced accuracy of $0.624$ was achieved on detecting landslides, by using also locations that the model has not been trained on. 
 \vspace{-0.35cm} 
\begin{figure}[!ht]
     \centering
     \includegraphics[scale=0.5]{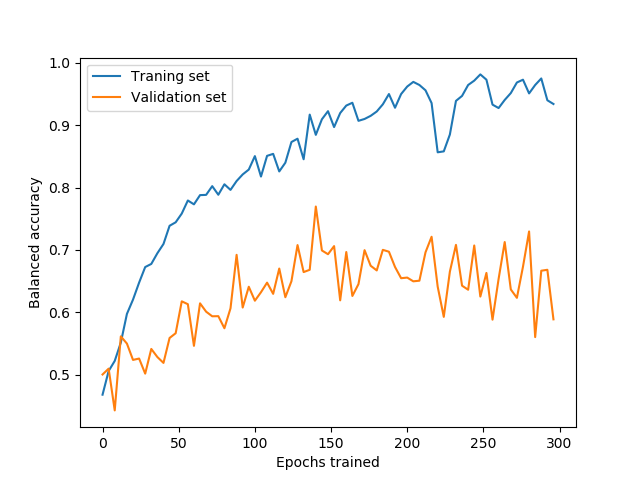}
     \caption{Training and Evaluation Accuracy }
     \label{fig:accuracies}
 \end{figure}
 \vspace{-0.35cm} 
As you can see in Figure \ref{fig:accuracies}, these accuracies are quite volatile and significantly higher evaluation accuracies were achieved at points during training. This is most likely due to the small size of our training and evaluation sets. It is important to note that our algorithm regardless is able to pick up significant signal after being trained on just a small sample of optical imagery. This is remarkable considering that typical Deep Learning algorithms require at least a few hundred examples from each class \cite{DL_scalability}.
A more large scale approach is likely to achieve good results. Since there are a lot of differences between distinct locations and images, significant generalization improvements could be achieved using additional data.
% \begin{figure}[!ht]
%      \centering
%      \includegraphics[scale=0.28]{images/landslides_type.png}
%      \caption{Landslide type }
%      \label{fig:landslides_type}
%  \end{figure}
 \vspace{-0.5cm} 
\section{Future developments}
The next step in the project development is to detect even medium and small size events. More images will be necessary. Another important parameter that must be considered is the image resolution. The final Sentinel 2 images have a resolution of $10\times 10 \ \meter $. A smaller resolution or further preprocessing may be necessary for accurate landslide prediction. 

While it is possible to detect landslides based on just optical imagery, it is likely that other types of data such as SAR from Sentinel-1 would be better suited for monitoring land movements. Some considerations have been already done in \cite{sicaullo},  where interferometry is used at this aim, and coherence of the images is discussed, as critical issue.  Sentinel-1 would allow also to overcome the problem related to  cloud coverage, proved to be a significant problem for Sentinel-2 data collection. Moreover, a system similar to that proposed in this paper, trained on SAR data or on a combination of SAR and optical data, might produce highly accurate results. The problem is that Sentinel-1 data have resulted not available on the period of interest, because of the USA shut down.
At this end, for future works Sentinel-1 data will be provided by the ESA centre for Earth observation (ESRIN) in Frascati, south of Rome, Italy. 

Another aspect to explore is the possibility to use the software for the classification of different types of past landslides. Landslide classification is primarily based on type of movement (falls, topples, slides, spreads, flows) and type of material (rock,
soil, mud and debris)  \cite{landslides_type}. A model similar to that proposed would likely succeed in this task, given sufficient amount of examples for each landslide type.

%\tiny
%\small
\footnotesize
\bibliographystyle{IEEEbib}
\bibliography{refs}
\end{document}